\documentclass[english,journal]{IEEEtran}
\usepackage[T1]{fontenc}
\usepackage[latin9]{inputenc}
\usepackage{array}
\usepackage{booktabs}
\usepackage{amsmath}
\usepackage{graphicx}
\usepackage{amssymb}

\begin{document}

\title{Linear Circuit Models for On-Chip Quantum Electrodynamics}

\author{Alp\'ar M\'aty\'as,~Christian Jirauschek,~\IEEEmembership{Member,~IEEE},~Federico
Peretti,~Paolo Lugli,~\IEEEmembership{Senior Member,~IEEE}, and~Gy\"orgy~Csaba
\linebreak(Dated: 17 June 2011, published as IEEE Trans. Microwave Theory Tech. 59, 65--71 (2011))
\linebreak \linebreak \copyright 2011 IEEE. Personal use of this material is permitted. Permission from IEEE must be obtained for all other uses, in any current or future media, including reprinting/republishing this material for advertising or promotional purposes, creating new collective works, for resale or redistribution to servers or lists, or reuse of any copyrighted component of this work in other works.
\thanks{The authors are with the Institute for Nanoelectronics, Technische Universit\"at M\"unchen,
Arcisstr. 21, D-80333 M\"unchen, Germany (e-mail: alparmat@mytum.de; jirauschek@tum.de; federico\_peretti2002@yahoo.it; lugli@tum.de).

G. Csaba is with the Institute for Nanoelectronics, Technische Universit\"at M\"unchen, D-80333 Munich, Germany, and also with the Center for Nano Science and Technology, University of Notre Dame, 46656 Notre Dame, USA  (e-mail: gcsaba@nd.edu).

Financial support was partially from the Emmy Noether program (DFG, JI115/1-1),
SFB 631 priority program of the DFG, the Erasmus programme of the European Union and 
the P\'azm\'any P\'eter Catholic University (ITK) of Budapest.}
}

\maketitle

\begin{abstract}
We present equivalent circuits that model the interaction of microwave resonators and quantum 
systems. The circuit models are derived from a general interaction Hamiltonian. 
Quantitative agreement between the simulated resonator transmission
frequency, qubit Lamb shift and experimental data will be shown. We demonstrate 
that simple circuit models, using only linear passive elements, can be very useful 
in understanding systems where a small quantum system is coupled to a classical microwave 
apparatus.

\end{abstract}
\IEEEpeerreviewmaketitle

\section{Introduction}

\label{intro}

\IEEEPARstart{C}{avity} quantum electrodynamics has long
been an active field to study the interaction of electromagnetic
radiation with matter, which is a fundamentally important topic in
physics~\cite{2002Sci...298.1372M,Berman01,Carmichael01}. Recently, cavity quantum electrodynamic experiments were
performed in the microwave regime~\cite{2004Natur.431..162W,2004PhRvA..69f2320B,2007Natur.445..515S,2008NatPh...4..686D},
where the cavity consisted of a high-quality factor ($Q$) microwave coplanar
resonator. The quantum system was a charge qubit built from
two Josephson junctions. Similar experiments were proposed to bring
other nanosystems (molecules \cite{2006NatPh...2..636A}, nanomagnets
\cite{2001Nanot..12..181T,Matyas01}, flux qubits~\cite{Matyas01,2007SuScT..20..814L,FahemLugli}) 
into interaction with coplanar waveguides. One of the most 
promising aspects of these experiments is that they
can provide circuits which integrate quantum mechanical behavior 
and 'conventional' (high-frequency) components on a chip.

It is well established how to construct circuit models from electromagnetic models for commonly 
used active or passive components, which obey 'classical' circuit theory. It has been also demonstrated
how one can build equivalent circuit models of basic two-state quantum
systems \cite{optical.csupor}. The goal of the present paper is to
show a systematic approach for building circuit models for an interacting
resonator-quantum system, using linear approximations. This can be useful 
for developing quantum systems coupled via a resonator \cite{2007Natur.449..443M} 
or modeling the large external circuitry coupled to the quantum system. The possibility
of this modeling approach was mentioned in \cite{2004PhRvA..69f2320B};
to our knowledge, though, our paper is the first to analyze and exploit
this model in detail. We investigate an experimental setup, similar
to the one in \cite{2008Sci...322.1357F}, composed of a high-Q, superconducting
microwave resonator, electrically coupled to a quantum circuit (often
referred as transmon-qubit \cite{2008PhRvL.101h0502H}).

During the past years the field of circuit quantum electrodynamics went through a considerable
growth. Many quantum phenomena known from optical and three dimensional microwave systems 
were also observed in circuit quantum electrodynamics and simulated, 
including systems of entangled quantum circuits~\cite{2007PhRvA..75c2329B} and 
resonators coupled via flux qubits~\cite{PhysRevB.81.144510}.
Such quantum electrodynamic (QED) systems need further optimization of both the quantum 
bit and the resonator, such as increasing qubit decay times and optimizing resonator losses~\cite{1131545}.

The natural application area of circuit models is the understanding
of the interaction between a relatively complex 'classical' circuitry
and a relatively simple quantum circuit. These models do not show
new physics, but rather facilitate the engineering of coupled microwave-quantum systems.

\section{Device geometry}
\label{sec:geom}
\begin{figure}[t]
 \centering \includegraphics[width=3.2in]{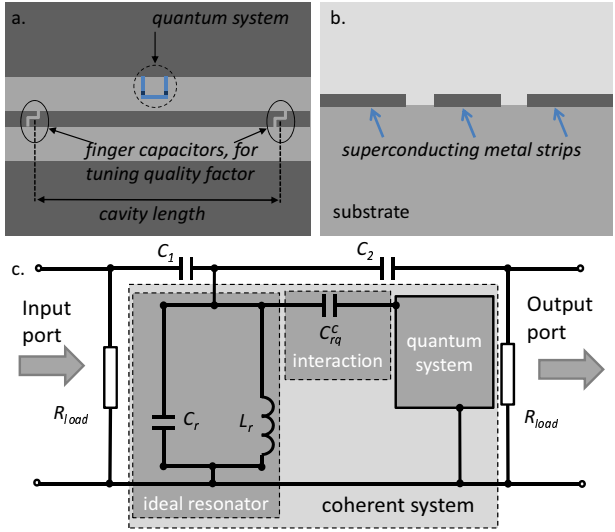}

\caption{ Geometry-sketch of the (a) longitudinal section of the resonator, showing the 
quantum system-loop that is placed near the field maximum and (b) the cross section of the investigated
coplanar waveguide resonator. The simplified equivalent circuit is shown in (c), 
representing the ideal resonator. Here, the  capacitors $C_{1}$ and $C_{2}$ correspond 
to the finger capacitors at the two ends of the resonator, and the load resistors are 
also included. The coupling to the quantum system is capacitive.}
\label{fig:7880_figure1} 
\end{figure}

The setup of a typical circuit quantum electrodynamic experiment is 
shown in Fig. \ref{fig:7880_figure1}. A main component is a superconducting 
coplanar waveguide resonator, displayed in Fig. \ref{fig:7880_figure1} (a) 
(longitudinal section) and (b) (cross section). A superconducting resonator 
can easily reach an unloaded quality factor in 
the range of $Q_{unloaded}\approx10^{6}$~\cite{2004Natur.431..162W,Frunzio01}. 
The loaded quality factor of the resonators is designed by using finger-type 
capacitors in the central strip. These typically determine the lifetime
of the resonator field (which should be high) and also the duration of
the measurement of the nanosystem (which should be low in order to avoid energy
relaxation and decoherence of the quantum system). The superconducting resonator 
is practically lossless and close to 
resonance, its port to port behavior can be modeled by the circuit elements 
in Fig. \ref{fig:7880_figure1} (c), where the LC 
parameters are:

\begin{eqnarray}
\begin{aligned}C_{r} & = & (Z_{0}\omega_{r})^{-1},\\
L_{r} & = & Z_{0}\omega_{r}^{-1}.\end{aligned}
\label{eq:lc}\end{eqnarray}

Here, $Z_{0}$ is the resonator impedance and $\omega_{r}$ is the resonator 
frequency. The 50\,$\Omega$ external environment 
is represented by the load resistors R$_{load}$ in Fig.~\ref{fig:7880_figure1} (c) which, 
together with the finger-type capacitors $C_{1}\approx C_{2}$ in the circuit model 
and the resonator frequency $\omega_{r}$, give the loaded quality factor of 
the resonator: $Q_{loaded}\approx 1/(R_{load}C_{1}\omega_{r}$).

The quantum circuit is placed in the gap of the resonator (see Fig. \ref{fig:7880_figure1} (a)) near the electric field 
maximum (magnetic field maximum) if it is an electrically (magnetically) interacting quantum system (i.e., charge qubit). 
Its presence slightly changes the values of the capacitance and inductance of the resonator and, thus, its 
resonance frequency. 

The total system can be divided into a resonator part, an interaction part and a quantum system part. 
The coupling in the case of electrically interacting systems is capacitive.

Throughout this paper we will formulate our models assuming electrically coupled systems. We will 
also, however, mention the parameters for the magnetic case in Section III. Instead of modeling 
different geometries, the resonator will be considered as a lumped circuit. For more detailed 
investigation of resonator geometries, the reader is referred to \cite{Matyas01,FahemLugli,2008PhRvL.101h0502H}. 

\section{Method }
\label{sec:Method used}

For the modeling of the interacting resonator-quantum system, we use a simplified 
(lossless) version of the circuit model in Fig. \ref{fig:7880_figure1} (c), and 
the coupling to the environment is neglected. The Hamiltonian is obtained by adding 
the energy stored in the ideal ($LC$) resonator, the energy stored in the interaction 
of the resonator with the quantum system, and the energy of the quantum system:

\begin{eqnarray}
 H_{tot}  = & \overbrace{\frac{1}{2}\left(C_{r}\hat{V}_{res}^{2}+L_{r}\hat{I}_{res}^{2}\right)}^\text{ideal res. energy} & +\nonumber \\
& + \overbrace{\frac{\hbar g}{2 V_{RMS}}\sigma_{1} \hat{V}_{res}}^\text{interaction energy} & +\overbrace{\frac{\hbar\omega_{q}}{2}\sigma_{3}}^\text{q. s. energy}.  \label{eq:intham}\end{eqnarray}
In the above equation, the different parts are illustrated by the boxes in Fig. \ref{fig:7880_figure1} (c). 

The resonator energy can be defined analogous to the classical energy of the $LC$ 
circuit in Fig. \ref{fig:7880_figure1} (c), however, the voltages and currents here 
are operators and their expectation value gives the voltage and current of the 
LC resonant circuit. The resonator root-mean square current ($I_{RMS}$) and voltage 
($V_{RMS}$) \cite{2004Natur.431..162W,2004PhRvA..69f2320B} is related to the minimum 
(or vacuum) field in the resonator. They are further discussed in Appendix \ref{derivRBE}. 

The interaction energy also contains the resonator voltage term ($\hat{V}_{res}$). 
Here, $g$ is related to the geometric capacitance of the resonator-quantum system 
$C^{c}_{rq}$, and $\sigma_{1}$ is related to the voltage of the quantum system. The 
$\sigma_{i}$, in the QED framework, are referred to as Pauli matrices and are the 
basic observables of the two state quantum system. The interaction term is a standard 
term used in quantum electrodynamics, and valid for every electrically interacting 
quantum system and resonator.
 
The last term in (\ref{eq:intham}), containing $\omega_{q}$ represents the energy 
level splitting ($\Delta E=\hbar \omega_{q}$) of the quantum system. This part can 
only be derived in terms of quantum mechanics.

For simulating the dynamics of the coupled system, we use the Liouville-von Neumann equation

\begin{equation}
\dot{A}=-\frac{j}{\hbar}\left[A,H_{tot}\right]\label{eq:neumann}\end{equation}
 with the Hamiltonian (\ref{eq:intham}). Here, A is an operator that represents either one of the Pauli matrices ($\sigma_{i}$) or the resonator voltage ($\hat{V}_{res}$) or current ($\hat{I}_{res}$) operator. Based on the von Neumann equation, we find the following coupled differential equation system:
\begin{eqnarray}
\dot{\lambda}_{1} & = & -\omega_{q}\lambda_{2}-\frac{\lambda_{1}}{T_{2}}\label{eq:rbe1}\\
\dot{\lambda}_{2} & = & \omega_{q}\lambda_{1}-g\lambda_{3}\frac{V_{res}}{V_{RMS}}-\frac{\lambda_{2}}{T_{2}}\label{eq:rbe2}\\
\dot{\lambda}_{3} & = & g\lambda_{2}\frac{V_{res}}{V_{RMS}}-\frac{\lambda_{3}+1}{T_{1}}\label{eq:rbe3}\\
\frac{\dot{V}_{res}}{V_{RMS}} & = & \omega_{r}\frac{I_{res}}{I_{RMS}}-\gamma\frac{V_{res}}{V_{RMS}}\label{eq:rbe4}\\
\frac{\dot{I}_{res}}{I_{RMS}} & = & -\omega_{r}\frac{V_{res}}{V_{RMS}}-g\lambda_{1}.\label{eq:rbe5}\end{eqnarray}
The above equations are a direct result of the von Neumann equation for the dissipationless case (if $T_{1}, T_{2} \rightarrow \infty $ and $\gamma \rightarrow 0$). We can observe that the first three equations (\ref{eq:rbe1}), (\ref{eq:rbe2}) and (\ref{eq:rbe3}) are the widely used Bloch equations \cite{book.opt1} written for a two-level quantum system; the last two are equations for the dynamics of normalized resonator voltage and currents, thus we will call Eqs. (\ref{eq:rbe1})-(\ref{eq:rbe5}) resonator-Bloch equations (RBEs). Dissipation is phenomenologically described by the decay and decoherence constants $T_{1}$ and $T_{2}$~\cite{book.opt1}, and the resonator loss, $1/\gamma$ is added similar as in \cite{JaynesCummingsOriginal}. The $\lambda_{i}$ variables are the expectation values of the Pauli matrices and represent the coherence vector characterizing the state of the quantum bit. The derivation of the resonator-Bloch equations is shown in Appendix \ref{derivRBE}.

The only approximation used here is to neglect the correlations between the qubit and field \cite{JaynesCummingsOriginal}, thus writing the expectation value of the normalized voltage and coherence vectors as a product of the individual expectation values:

\begin{equation}
\left\langle \sigma_{i}\hat{V}_{res}\right\rangle =\left\langle \sigma_{i}\right\rangle \left\langle \hat{V}_{res}\right\rangle =V_{res}\lambda_{i}.\label{eq:expval}\end{equation}

An equation system similar to the resonator-Bloch equations was introduced by Jaynes and Cummings \cite{JaynesCummingsOriginal,1962PhDT........41C}, written for the coupling of electric fields and dipole moments. We now use this model to find passive circuits describing the interaction of the  LC resonator-quantum system. 

\subsection{Modeling the system for large detuning}

\label{sec:large_detuning}

\begin{figure}[!t]
\centering\includegraphics[width=3.2in]{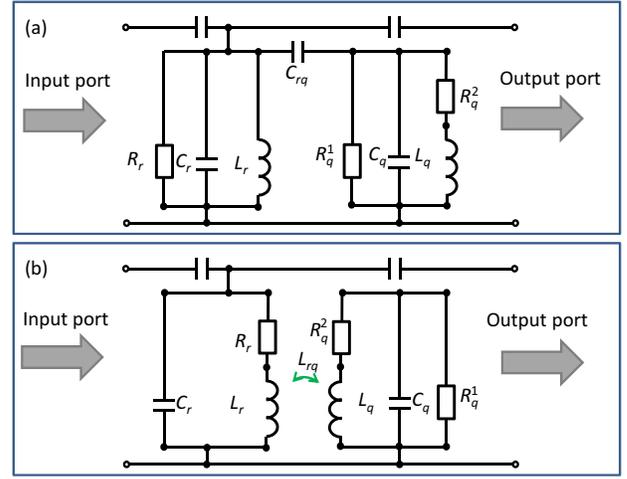}

\caption{Circuit models representing the linear quantum mechanical interaction
of the resonator and the qubit (i.e., $\lambda_{3}\approx\lambda_{3}^{0}$)
in case of (a) electric interaction (e.g., charge qubit) and (b) magnetic interactions (e.g., flux qubit). The coupling between the resonator and
quantum system is capacitive in (a) and inductive in (b). The quantum system behaves as a linear LC oscillator
that changes the frequency of the resonator. The outcoupling capacitors, corresponding to the resonator finger-type gaps, 
have very small values and thus can be added to the resonant circuits that model the linear interaction. }

\label{fig:7880_figure2} 
\end{figure}

In circuit quantum electrodynamic experiments, measurements are often performed in the so-called
dispersive regime, where the qubit-resonator detuning is much larger then the coupling frequency: 
$\Delta\gg g$. In this case, off-resonant interaction between the quantum system and the resonator does not switch the quantum system and $\lambda_{3}$ remains constant. Thus, we can neglect the oscillations of the inversion term in the resonator-Bloch equations ($\dot{\lambda}_{3}=0$) 
and perform a linear approximation by keeping $\lambda_{3}=\lambda_{3}^{0}$.
This is valid only if the measurement time is much smaller than the decay time
$T_{1}$. For a large decay time $T_{1}$ or short measurement time window, the system behaves linearly,
and the resonator-Bloch equations reduce to the four coupled linear differential equations
(we will call them linear RBEs): Eqs. (\ref{eq:rbe1}), (\ref{eq:rbe2}), (\ref{eq:rbe4})
and (\ref{eq:rbe5}) with $\lambda_{3}=\lambda_{3}^{0}$. These can
be modeled by the passive circuits shown in Fig.~\ref{fig:7880_figure2} and represent the interaction of (a) electric and (b) magnetic
quantum systems, with the resonator in the dispersive regime. Interaction
between the quantum system and resonator occurs through their coupling
capacitance in Fig.~\ref{fig:7880_figure2} (a) or mutual inductance in Fig.~\ref{fig:7880_figure2} (b), depending on the type of interaction (electric or magnetic). 

In the circuit models, the decoherence time from resonator-Bloch equations (\ref{eq:rbe1})
and (\ref{eq:rbe2}) is represented by the parallel resistor with 
$R_{q}^{1}C_{q}\approx T_{2}$; the series resistor, with 
$L_{q}/R_{q}^{2}\approx T_{2}$ and the resonator decay rate
$\gamma$, is given by $R_{r}C_{r}$ in case of electric and $L_{r}/R_{r}$
in case of magnetic interactions. The parameters of the circuits were
found as a direct result of the linear resonator-Bloch equations and are summarized in Table
\ref{MTT_table1}; for the derivation see Appendix \ref{derivcirc}. This table summarizes the central result of our work.

\begin{table*} [t]
\global\long\def\arraystretch{1.75}
 \centering \begin{tabular}{ccccccccc}
\toprule 
\multicolumn{1}{c}{Interaction} & $R_{q}^{1}$  & $R_{q}^{2}$  & $L_{q}$  & $C_{q}$  & $R_{r}$  & $L_{r}$  & $C_{r}$  & $C_{rq}/L_{rq}$\tabularnewline
\midrule 
Electric  & $\omega_{r}T_{2}(-\lambda_{3}^{0})Z_{0}$  & $\frac{(-\lambda_{3}^{0})Z_{0}\omega_{r}}{T_{2}\omega_{q}^{2}}$  & $\frac{(-\lambda_{3}^{0})Z_{0}\omega_{r}}{\omega_{q}^{2}}$  & $\frac{1}{Z_{0}\omega_{r}(-\lambda_{3}^{0})}-C_{rq}$  & $Z_{0}\frac{\omega_{q}}{\gamma}$  & $\frac{\omega_{q}Z_{0}}{\omega_{r}^{2}}$  & $\frac{1}{Z_{0}\omega_{q}}-C_{rq}$  & $\frac{g}{\omega_{r}\omega_{q}Z_{0}}$\tabularnewline
\midrule 
Magnetic  & $\frac{\omega_{q}^{2}T_{2}Z_{0}}{(-\lambda_{3}^{0})\omega_{r}}$  & $\frac{Z_{0}}{\omega_{r}T_{2}(-\lambda_{3}^{0})}$  & $\frac{Z_{0}}{\omega_{r}(-\lambda_{3}^{0})}-L_{rq}$  & $\frac{(-\lambda_{3}^{0})\omega_{r}}{Z_{0}\omega_{q}^{2}}$  & $Z_{0}\frac{\gamma}{\omega_{q}}$  & $\frac{Z_{0}}{\omega_{q}}-L_{rq}$  & $\frac{\omega_{q}}{Z_{0}\omega_{r}^{2}}$  & $\frac{gZ_{0}}{\omega_{r}\omega_{q}}$\tabularnewline
\bottomrule
\end{tabular}
\caption{Summary of the parameters calculated from the linear resonator-Bloch equations, if the
inversion is constant $(\lambda_{3}\approx\lambda_{3}^{0})$.}

\label{MTT_table1} 
\end{table*}

The physical parameters of the resonator are changed from the values
in (\ref{eq:lc}) to the values in Table \ref{MTT_table1}
due to the presence of the quantum system. The circuit models presented
in this section provide two resonance peaks in every case, which represent
the oscillations of the coupled system. The coupling between the quantum system and the 
resonator changes the electromagnetic response of the resonator in a way that depends 
on the state ($\lambda_{3}$) of the quantum system. The quantum state can be read out 
non-destructively from the measurement on the resonator. In physics this is often referred to 
as quantum non-demolition measurement.

\subsection{Modeling the system in its ground state}
\label{sec:ground_state}

Next, we will investigate the ground state of the coupled resonator-quantum system. 
We will also compare our simulation results mainly to measurements done in the ground 
state. In case of no applied field on the input port, the coupled resonator-quantum 
system will converge to the common ground state 
$[\lambda_{1},\lambda_{2},\lambda_{3},V_{res}/V_{RMS},I_{res}/I_{RMS}]=[0,0,-1,0,0]$,
due to the decay times $T_{1}$, $T_{2}$ and $1/\gamma$ in the resonator-Bloch equations. 
In the case of a small applied probe field on the resonator port, the system will oscillate 
around its ground state and the inversion remains unchanged (i.e., $\lambda_{3}\approx -1$). 

Rather than implementing the drive-fields in our resonator-Bloch equations, we start 
our simulations by taking a non-zero mode occupation of the resonator field. For this 
we take the initial condition: [0.04; 0; -0.999; 0; 0]. This corresponds to fields 
in the resonator-quantum system with an average photon number of approximately 0.01. 
Thus, for low fields in the resonator (approximately 0.01 photons \cite{2004PhRvA..69f2320B,2008Sci...322.1357F}), 
the nonlinear term $\lambda_{2}V_{res}/V_{RMS}$ in the resonator-Bloch equations will 
oscillate near $0$, as it is a product of two very small values. The inversion $\lambda_{3}$ 
will stay approximately constant and not change from its ground state even when the system 
has no detuning. As the inversion can be taken constant, the linear approximation done in 
the previous subsection holds, and we can take $\lambda_{3}^{0}\approx -1$.

In the next section, we will compare our derived resonator-Bloch equation model 
and circuit models to experimental data, and show the validity of our approximations. 

\section{Results }
\label{sec:results}

We have performed numerical and circuit simulations on a recently measured
resonator-transmon qubit device from \cite{2008Sci...322.1357F}. We took as input parameters:
resonator frequency $\omega_{r}/(2\pi)=6.44$~GHz, impedance $Z_{0}=50$~$\Omega$,
vacuum Rabi frequency $g/(2\pi)=266$~MHz and cavity decay rate $\gamma/(2\pi)=1.6$~MHz.
The dephasing time was approximately $T_{2}\approx1$~$\mathrm{\mu s}$,
which is known to be in good agreement with experimental findings
\cite{2008PhRvL.101h0502H}. Further, the linewidth of the qubit coherence
vector elements $\lambda_{1},\lambda_{2}$ is in good agreement with the
experiment (approximately 3~MHz) in \cite{2008Sci...322.1357F}. Simulations were performed 
in LTSPICE \cite{ltspice} based on the presented circuit models, using frequency domain AC analysis. 

The transmission of the circuit in Fig. \ref{fig:7880_figure2} (a) was calculated for 
the above parameters, setting $\lambda_{3}^{0}=-1$ as the system was in ground state. 
We investigated the dependence of the result with respect to quantum system frequency, 
while keeping the resonator frequency fixed. This was done by varying $\omega_{q}$ which 
was directly related to the parameters of our circuit model. 

\begin{figure}[!t]
\centering \includegraphics[width=3.2in]{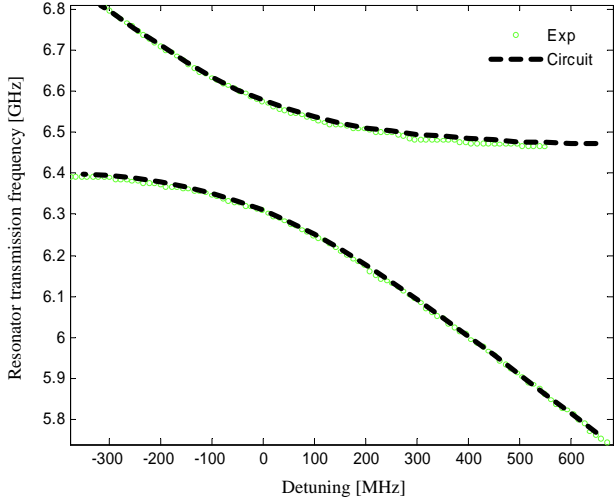}

\caption{Transmission frequency of the resonator as a function of detuning. 
Small circles represent experimental data from \cite{2008Sci...322.1357F} 
and the dashed lines represent our circuit model. The anticrossing behavior
is clearly visible and the separation is approximately given by the vacuum Rabi
frequency at $\Delta=0$. The excellent agreement confirms the
usability of our circuit model to calculate cavity transmission.}

\label{fig:7880_figure3} 
\end{figure}

The empty circles in Fig. \ref{fig:7880_figure3} represent data from~\cite{2008Sci...322.1357F} 
that has been transformed from flux to detuning coordinates so that our simulated results 
can be compared to the measured cavity transmission. We can see excellent agreement of 
the transmission frequency of our circuit (dashed line) with experimental data (small circles), 
for positive and negative detuning, showing the validity of the linear 
approximation and the excellent description of the anticrossing of the resonator-quantum 
system by the derived circuit model. At approximately $\Delta=0$, the peaks are separated 
by the vacuum Rabi frequency $g/(2\pi)$. The transmission graphs do not contain amplitude 
data, but we mention here, that the reduction of the peak far from the resonator frequency 
($\omega_{r}/(2\pi$)=6.44\,GHz) is also observed in our simulation. In the case of large 
detuning, the peaks are highly separated and the qubit peak gets highly reduced, while 
the resonator frequency peak approaches the original resonator frequency; this shows 
that for large detuning the two systems are decoupled.

\begin{figure}[!t]
\centering \includegraphics[width=3.2in]{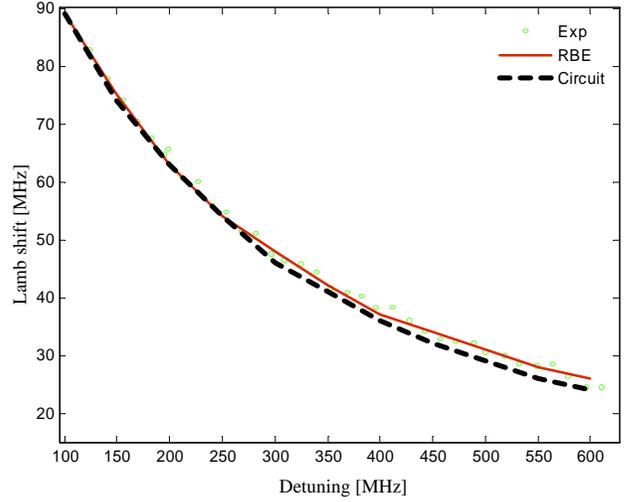}

\caption{Frequency (Lamb) shift of the quantum bit as a function of qubit detuning.
Small circles represent experimental data. Continuous lines represent results
extracted from the numerical solutions of the resonator-Bloch equations. Dashed lines indicate results
of our passive circuit model (Fig. \ref{fig:7880_figure2} (a))}

\label{fig:7880_figure4} 
\end{figure}

After successfully applying our circuit model to the description of the cavity transmission, 
we performed an analysis on the quantum bit frequency shift (Lamb shift). The cavity 
transmission was calculated as a function of the detuning; the Lamb shift was extracted 
by subtracting the bare cavity frequency and the detuning ($\Delta$) from the simulated 
qubit frequency. This was done by solving the full resonator-Bloch equations 
(\ref{eq:rbe1})-(\ref{eq:rbe5}) numerically using a basic Runge-Kutta solver 
and by employing our circuit models. The simulation results were compared to 
experimental data from \cite{2008Sci...322.1357F}, for a detuning ranging
from 100\,MHz to 600\,MHz. The comparison was done by changing the frequency 
of the qubit and keeping the resonator frequency constant. 

As shown in Fig. \ref{fig:7880_figure4}, we see an excellent agreement between the experimental 
Lamb shift (small empty circles) and the numerical solutions of the resonator-Bloch equations 
(solid line) as well as our linear circuit model (dashed line). Thus, we can affirm that also 
the frequency shift of the quantum system (Lamb shift) is well described by our circuit model. 
To this point we only did simulations of the quantum system in its ground state. For the 
numerical solution of the full resonator-Bloch equations, we used the initial conditions 
$[\lambda_{1},\lambda_{2},\lambda_{3},V_{res}/V_{RMS},I_{res}/I_{RMS}]=[0.04,0,-0.999,0,0]$ 
as discussed in Section \ref{sec:ground_state}. For the circuit model, we set $\lambda_{3}^{0}=-1$. 

We will now investigate the case, when the quantum system is away from its ground state, and 
its frequency is far detuned from the resonator frequency. In this case one can perform a 
non-demolition readout of the quantum state, i.e., read out the inversion without changing it. 
This can be done if the detuning tends to infinity (practically is large enough so that 
the inversion changes only slightly). We then know from Rabi's solutions that the 
inversion will not change (i.e., the resonator is unable to switch the quantum circuit during measurement). 
For an optimal quantum state  readout, however, the state of the quantum system should not decay 
during measurement. Based on our circuit model and resonator-Bloch equations presented earlier, 
we now estimate the value of the decay time $T_{1}$ for an optimal readout. 

\begin{figure}[t]
\centering \includegraphics[width=3.2in]{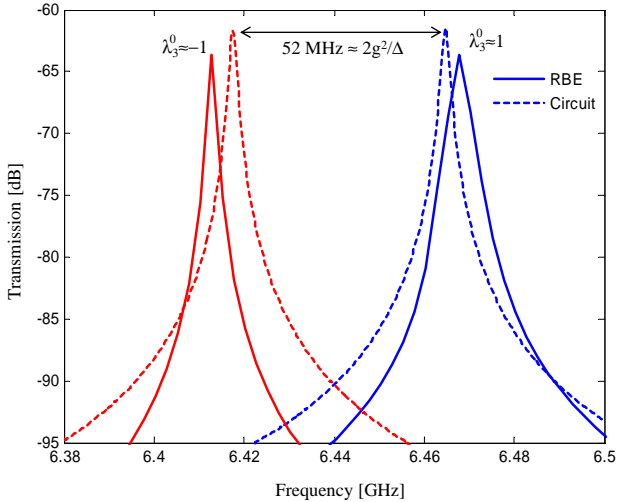}

\caption{Theoretical demonstration of our circuit model showing the state dependent resonator frequency shift, 
that opens the possibility of a quantum state measurement. The quantum system decay time was  20\,$\mu s$, that was large enough 
not to influence the linewidth of the excited state. The shift is approximately $g^{2}/\Delta$. The slight disagreement 
of the peak positions results from the Fourier transform on the finite simulation window of the output voltage signal 
calculated with the resonator-Bloch equations.}

\label{fig:7880_figure5} 
\end{figure}

The quantum state readout measurement is performed in the dispersive regime, as discussed in 
Section \ref{sec:large_detuning}. In the case of small decay times $T_{1}$, if the quantum system is in its excited state, it will quickly relax to the 
ground state; this limits the measurement time and, thus, the Fourier window of our numerical simulations. On the other hand, for large decay times, the 
quantum-state dependent cavity shift can be simulated, as shown in Fig. \ref{fig:7880_figure5}. The simulated cavity-pull linewidths extracted from the Fourier 
transform of the numerical resonator-Bloch equation solutions and from the frequency domain simulation of our circuit model agree well. This shows that the 
broadening of the peaks is mainly due to decoherence $T_{2}$ and cavity decay ($1/\gamma$) and not the qubit decay $T_{1}$, which was not included in the 
circuit model. In this limit, the quantum-state dependent cavity pull is resolvable, and we find that for this case $T_{1}\geq20$~$\mu s$ (or more) 
would be required, so that the two linewidths approximately agree. For smaller decay times, the numerical simulations of the resonator-Bloch equations show 
high asymmetry of the two peaks due to the non-Lorentzian behavior of the spectral line corresponding to the excited qubit state ($\lambda_{3}^{0}=1$); 
this is due to the decay of the state during simulation. In order to avoid the decay of the quantum state the simulation window can be reduced. For a too 
small simulation window, however, the Fourier transform of the signal becomes inaccurate. 

\section{Conclusions}
\label{sec:concl}

We have derived circuit models for the understanding and modeling
of superconducting coplanar waveguide resonators interacting with quantum
systems. They can easily be used to model the cavity transmission, Lamb shift and quantum non-demolition measurement.
We find their application straightforward in understanding experimental data and 
estimating decay times for an optimal quantum state readout. Our models can be extended, e.g., 
by adding further circuits to represent additional qubits.
Also, embedding quantum systems in large-scale classical circuitry is straightforward in our circuit models.

\appendices

\section{Derivation of the resonator-Bloch equations}
\label{derivRBE} 

For the derivation of the resonator-Bloch equations, we need to review a few concepts 
widely used and well established in quantum optics. The starting point is to define the resonator voltage, similar as in \cite{2004Natur.431..162W,2004PhRvA..69f2320B}. 
The voltages and currents on the $LC$ circuit are defined as:

\begin{equation}
V_{RMS}^{-1}\hat{V}_{res}=a+a^{+}\label{eq:Vres},\end{equation}
\begin{equation}
I_{RMS}^{-1}\hat{I}_{res}=-j(a-a^{+})\label{eq:Ires}.\end{equation}
Here the terms $a^{+}$ and $a$ represent the emission and absorption operators. The commutator properties of these operators are

\begin{align} 
[a,a^{+}]=1, [a,a^{+}a]=a, [a^{+},a^{+}a]=-a^{+}, \nonumber \\
[a,a]=[a^{+},a^{+}]=[a^{+}a,a^{+}a]=0. \label{eq:ARBE5p}
\end{align}
The brackets represent a commutator relation (also used in the von Neumann equation (\ref{eq:neumann})). 

The normalization terms $V_{RMS}$ and $I_{RMS}$ represent the root-mean square voltage  and 
current~\cite{2004Natur.431..162W,2004PhRvA..69f2320B}, on the resonator:

\begin{align} 
V_{RMS}=\sqrt{\hbar\omega_{r}/(2C_{r})}, \\
I_{RMS}=\sqrt{\hbar\omega_{r}/(2L_{r})}. \label{eq:ARMS}
\end{align}

If we substitute the voltage and current operators ((\ref{eq:Vres}) and (\ref{eq:Ires})) into the 
general Hamiltonian shown in (\ref{eq:intham}) and use the commutator operations defined in (\ref{eq:ARBE5p}), we can get the general Hamiltonian

\begin{eqnarray}
\begin{aligned}H_{tot} & =\overbrace{\hbar\omega_{r}a^{+}a}^\text{ideal
res. energy}+\overbrace{\frac{\hbar g}{2}\sigma_{1}(a+a^{+})}^\text{interaction energy}+\overbrace
{\frac{\hbar\omega_{q}}{2}\sigma_{3}}^\text{q. s. energy},\end{aligned}
\label{eq:intham1}\end{eqnarray}
widely used in cavity quantum electrodynamics \cite{2004PhRvA..69f2320B,book.opt1,2005PhRvL..95u3001S,book.opt2}. 

For each Pauli matrix $\sigma_{i}, (i=1, 2, 3)$ and for the normalized current and voltage 
in (\ref{eq:Vres}) and (\ref{eq:Ires}), we used the von Neumann equation (\ref{eq:neumann}) for 
finding the resonator-Bloch equations. 

We now show the derivation of the dynamic equation for the resonator normalized current 
(\ref{eq:rbe5}). The other resonator-Bloch equations (for coherence vector and resonator 
normalized voltage) can be derived analogously. First, the normalized current can be written as
\begin{align}
I_{RMS}^{-1}\partial_{t}\hat{I}_{res}=-j\hbar^{-1}\left[I_{RMS}^{-1}\hat{I}_{res},H_{tot}\right]= \nonumber \\
-j\hbar^{-1}I_{RMS}^{-1}\left(\hat{I}_{res}H_{tot}-H_{tot}\hat{I}_{res}\right). \label{eq:ARBE5} 
\end{align}
Using the properties of the creation/annihilation operators from (\ref{eq:ARBE5p}), the above  von Neumann equation simplifies to
\begin{equation}
I_{RMS}^{-1}\partial_{t}\hat{I}_{res}=-\omega_{r}(a+a^{+})-g\sigma_{1}.\label{eq:ARBE5d1}\end{equation}
Using the definition in (\ref{eq:Vres}) and
taking the expectation values of the time dependent variables in the above equation, we get (\ref{eq:rbe5}).

\section{Derivation of the equivalent linear circuit}
\label{derivcirc}

If the inversion is constant ($\lambda_{3}=\lambda_{3}^{0}$) and the system does not decay (i.e., $T_{1}\rightarrow \infty$), 
we can linearize the resonator-Bloch equations and thus get an equivalent circuit. In this case (\ref{eq:rbe3}) is automatically fulfilled, 
and the other resonator-Bloch equations (\ref{eq:rbe1}), (\ref{eq:rbe2}), (\ref{eq:rbe4}) and (\ref{eq:rbe5}) simplify to the linear differential equation system:
\begin{eqnarray}
\dot{\lambda}_{1} & = & -\omega_{q}\lambda_{2}-T_{2}^{-1}\lambda_{1}\label{eq:lrbe1}\\
\dot{\lambda}_{2} & = & \omega_{q}\lambda_{1}-g\lambda_{3}^{0}V_{RMS}^{-1}V_{res}-T_{2}^{-1}\lambda_{2}\label{eq:lrbe2}\\
V_{RMS}^{-1}\dot{V}_{res} & = & I_{RMS}^{-1}\omega_{r}I_{res}-\gamma V_{RMS}^{-1}V_{res}\label{eq:lrbe4}\\
I_{RMS}^{-1}\dot{I}_{res} & = & -V_{RMS}^{-1}\omega_{r}V_{res}-g\lambda_{1}.\label{eq:lrbe5}
\end{eqnarray}
By rewriting the above as two second order equations, we get:
\begin{eqnarray}
\ddot{\lambda}_{2} &=& -\omega_{q}^{2} \lambda_{2}-T_{2}^{-1}\omega_{q}\lambda_{1} - g\lambda_{3}^{0}V_{RMS}^{-1}\dot{V}_{res} \nonumber \\
& &-T_{2}^{-1}\dot{\lambda}_{2} \label{eq:so1}\\
I_{RMS}^{-1}\ddot{I}_{res} &=&  V_{RMS}^{-1}\omega_{r} \gamma V_{res}-I_{RMS}^{-1}\omega_{r}^{2} I_{res} -g \dot{\lambda}_{1}. \,\,\,\,\,\, \label{eq:so2}
\end{eqnarray}
The above equations can be equivalently found by using:
\begin{eqnarray}
\dot{\lambda}_{1} & = & -\omega_{q}\lambda_{2}-\omega_{q}^{-1}V_{RMS}^{-1}g\lambda_{3}^{0}\dot{V}_{res}-T_{2}^{-1}\lambda_{1}\label{eq:lrbe11}\\
\dot{\lambda}_{2} & = & \omega_{q}\lambda_{1}-T_{2}^{-1}\lambda_{2}\label{eq:lrbe21}\\
\dot{V}_{res} & = & \omega_{r}Z_{0}I_{res}+\omega_{r}^{-1}gV_{RMS}\dot{\lambda}_{1}-\gamma V_{res}\label{eq:lrbe41}\\
\dot{I}_{res} & = & -Z_{0}^{-1}\omega_{r}V_{res}.\label{eq:lrbe51}
\end{eqnarray}
Introducing the quantum system voltage and current as:
\begin{eqnarray}
V_{q}=V_{RMS}\lambda_{1},\\
I_{q}=(-\lambda_{3}^{0})^{-1}\omega_{r}^{-1}I_{RMS}\omega_{q}\lambda_{2},
\end{eqnarray}
(\ref{eq:lrbe11})-(\ref{eq:lrbe51}) change into equations for two 
capacitively coupled resonant circuits, with  parameters shown in Table \ref{MTT_table1}.

\section*{Acknowledgment}

The authors are grateful to Prof. \'A. Csurgay, Faculty of Information Technology, P\'azm\'any P\'eter Catholic University, Budapest, Hungary, and Prof. W.
Porod, University of Notre Dame, Notre Dame, IN, for initiating their research on circuit models of quantum systems.
The work on resonators started in collaboration with Prof. R. Gross, Walther Meissner Institute, Garching, Germany.

 \bibliographystyle{IEEEtran} 

\begin{IEEEbiographynophoto}
{Alp\'ar M\'aty\'as} Alp\'ar M\'aty\'as was born in Aiud, Romania, in 1983.
He received his M.Sc. degrees in electrical engineering from the P\'azm\'any P\'eter
Catholic university of Budapest, in 2007.

Since 2007 he is a Ph.D. student at the TU M\"{u}nchen in Germany. His main research interests
are modeling and optimizing quantum cascade lasers in the THz and infrared regimes and 
simulations including the interaction between resonators and quantum systems.
\end{IEEEbiographynophoto}    
\begin{IEEEbiographynophoto}
{Christian Jirauschek} Christian Jirauschek was born in Karlsruhe, Germany,
in 1974. He received his Dipl-Ing. and doctoral degrees in electrical
engineering in 2000 and 2004, respectively, from the Universit\"{a}t Karlsruhe
(TH), Germany.

From 2002 to 2005, he was a Visiting Scientist at the
Massachusetts Institute of Technology (MIT), Cambridge, MA. Since 2005, he has
been with the TU M\"{u}nchen in Germany, first as a Postdoctoral Fellow and
since 2007 as the Head of an Independent Junior Research Group (Emmy-Noether
Program of the DFG). His research interests include modeling in the areas of
optics and device physics, especially the simulation of quantum devices and
mode-locked laser theory.

Dr. Jirauschek is a member of the IEEE, the German
Physical Society (DPG), and the Optical Society of America. Between 1997 and
2000, he held a scholarship from the German National Merit Foundation
(Studienstiftung des Deutschen Volkes).
\end{IEEEbiographynophoto} 
\begin{IEEEbiographynophoto}{Federico Peretti} Federico Peretti was born in Fermo, Italy in 1980.
He received his degree in electronic engineering and doctoral degree in electromagnetism and bioengineering 
in 2005 and 2009, respectively, from the Universit\`a Politecnica Delle Marche, Italy.
From 2006 to 2009 he was Scientist at the Institute for Nanoelectronics at the Technische Universit\"at M\"unchen, 
where his interests included modeling in the areas of microwaves and of quantum devices, especially the simulation 
of coplanar resonators for interactions with two-level systems.
\end{IEEEbiographynophoto}
\begin{IEEEbiographynophoto}
{Paolo Lugli} Paolo Lugli received the Laurea degree in physics at the University of Modena, 
Italy, in 1979. In 1981 he joined Colorado State University, Fort Collins, CO, 
where he received his Master of Science in 1982 and his Ph.D. in 1985, both in 
Electrical Engineering. In 1985 he joined the Physics Department of the University 
of Modena as Research Associate. From 1988 to 1993 he was Associate Professor of 
``Solid State Physics'' at the ``Engineering Faculty'' of the 2nd University of Rome ``Tor Vergata''. 
In 1993 he was appointed as Full Professor of ``Optoelectronics'' at the same university. 
In 2003 he joined the Technische Universit\"at M\"unchen where he was appointed head of the 
newly created Institute for Nanoelectronics.He is author of more than 250 scientific 
papers and co-author of the books ``The Monte Carlo Modelling for Semiconductor Device Simulations'' 
(Springer, 1989) and ``High Speed Optical Communications'' (Kluver Academic, 1999). 
In 2004, he served as General Chairman of the IEEE International Conference on Nanotechnology held in Munich.

His current research interests involve the modeling, fabrication and characterization 
of organic devices for electronics and optoelectronics applications, the design of 
organic circuits, the numerical simulation of microwave  semiconductor devices, and 
the theoretical study of transport processes in nanostructures. 
\end{IEEEbiographynophoto} 
\begin{IEEEbiographynophoto}
{Gy\"orgy Csaba} Gy\"orgy Csaba was born in Budapest, Hungary, in 1974. He 
received the M.Sc. degree from the Technical  University of Budapest in 1998 
and his PhD degree from the University of Notre Dame in 2003. From 2004 to 2009 he worked as a Research 
Assistant at the Technische Universit\"at M\"unchen, Germany and in 2010 he joined the 
faculty of  the university of Notre Dame as a research assistant professor.
His research interests are in circuit-level modeling of nanoscale systems (especially 
magnetic devices) and exploring their applications for nonconventional architectures, 
such as magnetic computing and physical cryptography.
\end{IEEEbiographynophoto}

\end{document}